\documentclass[prl,aps,showpacs,epsf, twocolumn, superscriptaddress]{revtex4-1}

\usepackage{graphics}
\usepackage{amsfonts}
\usepackage{amsmath}
\usepackage{epsfig}

\newif\ifdraft \drafttrue

\catcode`\ä = \active \catcode`\ö = \active \catcode`\ü = \active
\catcode`\Ä = \active \catcode`\Ö = \active \catcode`\Ü = \active
\catcode`\ß = \active \catcode`\é = \active \catcode`\è = \active
\catcode`\ë = \active \catcode`\ô = \active \catcode`\ê = \active
\catcode`\ø = \active \catcode`\ò = \active \catcode`\í = \active
\catcode`\Ó = \active \catcode`\ú = \active \catcode`\á = \active
\catcode`\ã = \active \catcode`\à = \active
\defä{\"a} \defö{\"o} \defü{\"u} \defÄ{\"A} \defÖ{\"O} \defÜ{\"U} \defß{\ss} \defé{\'{e}}
\defè{\`{e}} \defë{\"{e}} \defô{\^{o}} \defê{\^{e}} \defø{\o} \defò{\`{o}} \defí{\'{i}}
\defÓ{\'{O}} \defú{\'{u}} \defá{\'{a}} \defã{\~{a}}\defà{\`{a}}

\begin{document}

\title{Feynman diagrams versus Fermi-gas Feynman emulator}

%

\author{K. Van Houcke}
\address{Department of Physics, University of Massachusetts, Amherst, MA 01003, USA}
\address{Department of Physics and Astronomy, Ghent University,
Proeftuinstraat 86, B-9000 Ghent, Belgium}
\author{F. Werner}
\address{Department of Physics, University of Massachusetts, Amherst, MA 01003, USA}
\address{ Laboratoire Kastler Brossel, Ecole Normale Sup\'{e}rieure, UPMC-Paris 6, CNRS,
24 rue Lhomond, 75005 Paris, France}
\author{E. Kozik}
\address{Theoretische Physik, ETH Z\"urich, CH-8093 Z\"urich}
\address{Centre de Physique Th\'eorique, Ecole Polytechnique, 91128 Palaiseau Cedex, France}
\author{N. Prokof'ev}
\address{Department of Physics, University of Massachusetts, Amherst, MA 01003, USA}
\address{Russian Research Center ``Kurchatov Institute", 123182 Moscow, Russia}
\author{B. Svistunov} 
\address{Department of Physics, University of Massachusetts, Amherst, MA 01003, USA}
\address{Russian Research Center ``Kurchatov Institute", 123182 Moscow, Russia}
\author{M.~J.~H.~Ku}
\address{Department of Physics, MIT-Harvard Center for Ultracold Atoms, and Research Laboratory of Electronics,
\\ MIT, Cambridge, Massachusetts 02139, USA}
\author{A.~T.~Sommer}
\address{Department of Physics, MIT-Harvard Center for Ultracold Atoms, and Research Laboratory of Electronics,
\\ MIT, Cambridge, Massachusetts 02139, USA}
\author{L. W. Cheuk}
\address{Department of Physics, MIT-Harvard Center for Ultracold Atoms, and Research Laboratory of Electronics,
\\ MIT, Cambridge, Massachusetts 02139, USA}
\author{A. Schirotzek}
\address{Advanced Light Source, Lawrence Berkeley National Laboratory, Berkeley, California 94720, USA}
\author{M. W. Zwierlein}
\address{Department of Physics, MIT-Harvard Center for Ultracold Atoms, and Research Laboratory of Electronics,
\\ MIT, Cambridge, Massachusetts 02139, USA}




\newcommand{\li}{$^6$Li}
\newcommand{\na}{$^{23}$Na}
\newcommand{\cs}{$^{133}$Cs}
\newcommand{\kk}{$^{40}$K}
\newcommand{\rb}{$^{87}$Rb}
\newcommand{\vect}[1]{\mathbf #1}
\newcommand{\g}{g^{(2)}}
\newcommand{\one}{$\left|\uparrow\right>$}
\newcommand{\two}{$\left|\downarrow\right>$}
\newcommand{\V}{V_{12}}
\newcommand{\GammaSD}{\ensuremath{\Gamma_{\mathrm{sd}}}}
\newcommand{\GammaColl}{\ensuremath{\Gamma_{\mathrm{coll}}}}
\newcommand{\normSD}{\ensuremath{\tilde{\Gamma}_{\mathrm{sd}}}}
\newcommand{\normDs}{\ensuremath{\tilde{D}_{\mathrm{s}}}}
\newcommand{\Ds}{\ensuremath{D_{\mathrm{s}}}}
\newcommand{\chis}{\ensuremath{\chi_{\mathrm{s}}}}
\newcommand{\normCs}{\ensuremath{\tilde{\chi}_{\mathrm{s}}}}
\newcommand{\ag}{\ensuremath{\alpha_{\mathrm{\Gamma}}}}
\newcommand{\ad}{\ensuremath{\alpha_{\mathrm{D}}}}
\newcommand{\kfa}{\frac{1}{k_F a}}
\newcommand{\resField}{\ensuremath{834 \,\rm G}}
\newcommand{\kickField}{\ensuremath{50 \,\rm G}}
\newcommand{\prepField}{\ensuremath{300 \,\rm G}}
\newcommand{\zup}{\ensuremath{z_\uparrow}}
\newcommand{\zdown}{\ensuremath{z_\downarrow}}
\newcommand{\vzup}{\ensuremath{v_{z\uparrow}}}
\newcommand{\vzdown}{\ensuremath{v_{z\downarrow}}}

\begin{abstract}
Precise understanding of strongly interacting fermions, from electrons in modern materials to nuclear matter, presents a major goal in modern physics. However, the theoretical description of interacting Fermi systems is usually plagued by the intricate quantum statistics at play. Here we present a cross-validation between a new theoretical approach, Bold Diagrammatic Monte Carlo (BDMC), and precision experiments on ultra-cold atoms. Specifically, we compute and measure with unprecedented accuracy the normal-state equation of state of the unitary gas, a prototypical example of a strongly correlated fermionic system. Excellent agreement demonstrates that a series of Feynman diagrams can be controllably resummed in a non-perturbative regime using BDMC. This opens the door to the solution of some of the most challenging problems across many areas of physics.
\end{abstract}

\maketitle

In his seminal 1981
lecture~\cite{feyn82quantumsimulator}, Feynman argued that an arbitrary quantum system cannot be efficiently simulated with a classical universal computer, because generally,
quantum statistics can only be imitated with a classical theory if probabilities are replaced with negative (or complex) weighting factors.
For the majority of many-particle models this indeed leads to the so-called sign problem 
which has remained an insurmountable obstacle.
According to Feynman, the only way out is to employ computers
made out of quantum-mechanical elements~\cite{feyn82quantumsimulator}.
The recent experimental breakthroughs in cooling, probing and controlling strongly interacting quantum gases prompted a challenging effort to use this new form of quantum matter to realize Feynman's emulators of fundamental microscopic models~\cite{feyn82quantumsimulator,bloc08review}.
Somewhat ironically,
Feynman's arguments which led him to the idea of emulators may be defied by
a theoretical method that he himself devised, namely Feynman diagrams.
This technique organizes the calculation of a given physical quantity as a series of diagrams 
representing all possible ways particles can propagate and interact (see, e.g., Ref.~\cite{fett71}).
For the many-body problem, this diagrammatic expansion is commonly
used either in perturbative regimes or within uncontrolled
approximations. However, the introduction of Diagrammatic Monte
Carlo recently allowed to go well beyond the first few diagrams, 
and even reach convergence of the series in a moderately correlated regime~\cite{vanho08,kozi10hub}.


 In this Letter we show that for a strongly correlated system and down to a phase transition, the diagrammatic series
can still be given a mathematical meaning and leads to controllable results within
Bold Diagrammatic Monte Carlo.
This approach, proposed in Refs.~\cite{prok07signproblem, prok08boldpolaron,vanho08}, is first implemented here for the many-body problem.
We focus on the unitary gas, i.e. spin-1/2
fermions with zero-range interactions at infinite scattering length~\cite{ingu08varenna,gior08review,zwer11book}. 
This system offers
the unique possibility to
stringently test our theory against a quantum emulator realized here
 with trapped ultracold $^6{\rm Li}$ atoms at a broad Feshbach resonance~\cite{ingu08varenna,gior08review,zwer11book}.
This experimental validation is indispensable for our theory based on resummation of a possibly divergent series:
although the physical answer is shown to be independent from the applied resummation technique -- suggesting that the procedure is adequate -- its mathematical validity remains to be proven.
In essence, Nature provides the `proof'. This presents the first -- though long-anticipated -- compelling example of
how ultra-cold atoms can guide new microscopic theories for strongly
interacting quantum matter. 

At unitarity,
the disappearance of an interaction-imposed length scale
leads to scale invariance.
This property renders the model relevant for other physical systems such as neutron matter.
It also makes the balanced (i.e., spin-unpolarised) unitary gas ideally suited for the experimental high-precision determination of the equation of state (EOS) described below.
Finally, it implies the absence of a small parameter, making the problem notoriously difficult to solve.

In traditional Monte Carlo approaches,
which simulate a finite piece of matter,
the sign problem
causes an exponential increase of the computing time with system size and inverse temperature.
In contrast, BDMC simulates a
mathematical answer in the thermodynamic limit. 
This radically changes the role
of the fermionic sign.
Diagrammatic contributions are sign-alternating with order,
topology and values of internal variables.
Since the number of graphs grows factorially with diagram order,
a near cancellation between these contributions is actually necessary for the series to be resummable by techniques requiring a finite radius of convergence.
We find that this `sign blessing' indeed takes place.

\begin{figure}
    \begin{center}
    \includegraphics[angle=0, width=\columnwidth]{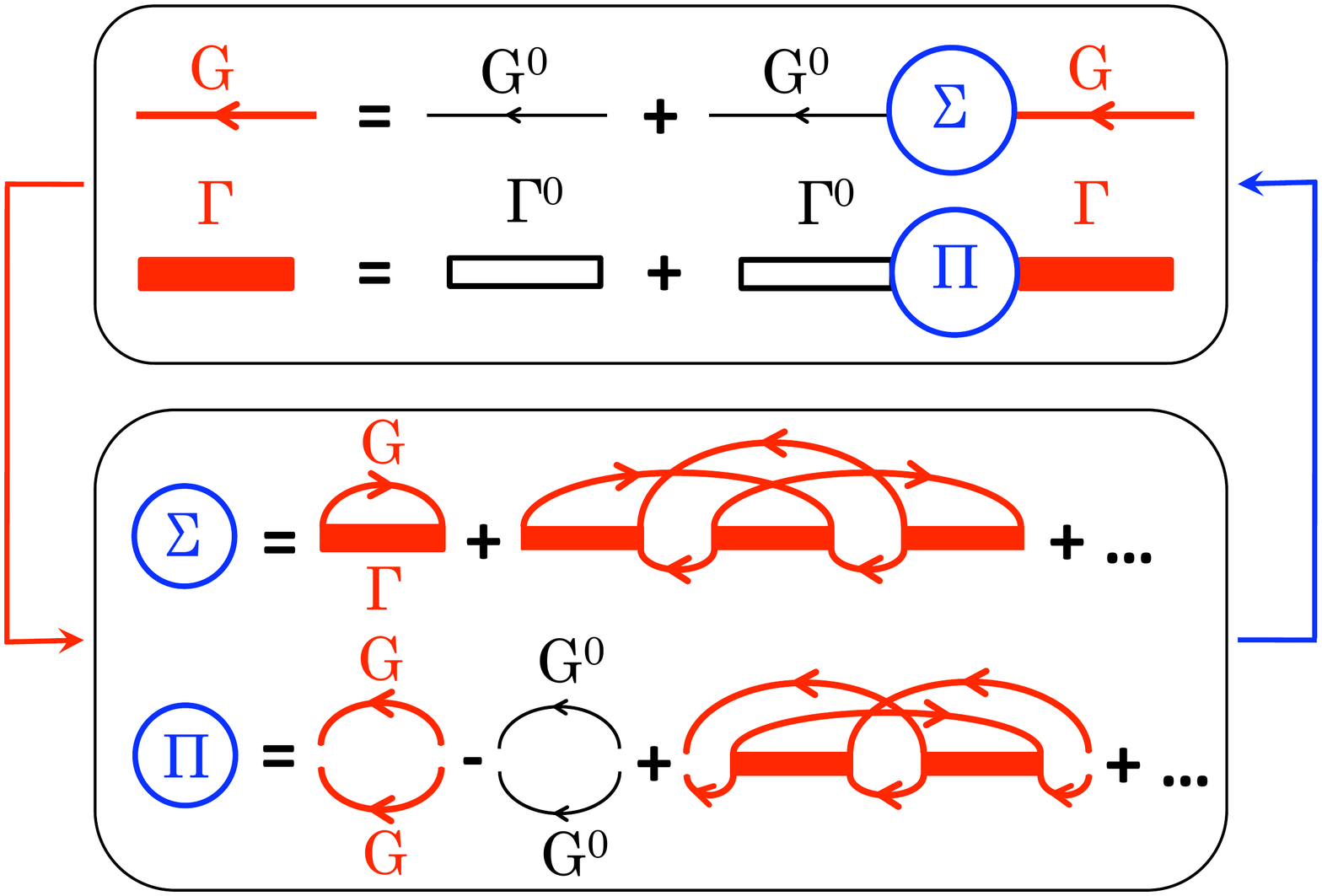}
    \caption[Title]{{\bf Bold Diagrammatic Monte Carlo}
     evaluates skeleton diagrammatic series for the self-energy $\Sigma$ and the
pair self-energy $\Pi$ (lower box). The diagrams are built on dressed
one-body propagators $G$
and pair propagators $\Gamma$, which themselves are the solution of the Dyson and Bethe-Salpeter equations (upper box). 
This cycle is repeated until convergence is reached.
$G^0$ is the non-interacting propagator, and
$\Gamma^0$ is the partially dressed
pair propagator obtained by summing the bare ladder diagrams.
}
    \label{fig:diagram}
    \end{center}
\end{figure}


In essence, BDMC solves the full quantum many-body problem by stochastically summing
all the skeleton diagrams for irreducible single-particle self-energy $\Sigma$ 
and pair self-energy $\Pi$, 
expressed in terms of bold (i.e., fully dressed)
single-particle and pair propagators $G$ and $\Gamma$
which are determined self-consistently (see Fig.~1).
The density EOS (i.e., the relation between total density $n$,  chemical potential $\mu$ and temperature $T$) 
is given by $G$ at zero distance and imaginary time, $n(\mu,T)=2\,G(r=0,\tau=0^-)$.
The thermodynamic limit can be taken analytically. 
The sum of ladder diagrams built on the bare single-particle propagator 
defines a partially dressed pair propagator $\Gamma^0$.
Since $\Gamma^0$ is well defined for the zero-range continuous-space interaction,
the zero-range limit can also be taken analytically.
This is in sharp contrast with other numerical methods~\cite{bulg06TC,buro06TC,goul10tc} where taking the thermodynamic and zero-range limits is computationally very expensive.
BDMC performs a random walk in the space of irreducible diagrams using local updates.
The simulation is run in a self-consistent cycle (along the lines 
of Ref.~\cite{prok07signproblem}) until convergence is reached. Full details will
be presented elsewhere. In essence, our approach upgrades the standard many-body theories
based on one lowest-order diagram
(see, e.g., Refs.~\cite{
stri11lect,haus94}) to millions
of graphs.

In the quantum degenerate regime,
 we do not observe convergence
of the diagrammatic series for $\Sigma$ and $\Pi$ evaluated up to order $9$.
Here, order $N$ means 
$\Sigma$-diagrams with $N$ vertices (i.e., $N$ $\Gamma$-lines) and $\Pi$-diagrams with $N-1$ vertices.
To extract the infinite-order result, we apply the following Abelian resummation methods~\cite{Hardy49}.
The contribution of all diagrams of order $N$ is multiplied by $e^{- \epsilon \lambda_{N-1}}$
where $\lambda_n$ depends on the resummation method:
(i) $\lambda_n=n \, \log n$ (with $\lambda_0=0$) for Lindel\"of~\cite{Hardy49},
(ii)
$\lambda_n=(n-1) \, \log (n-1)$ (with $\lambda_0=\lambda_1=0$) for ``shifted Lindel\"of'',
or (iii)
$\lambda_n=n^2$ for  Gaussian~\cite{fru92resum}.
A full 
simulation is performed for each $\epsilon$, and
the final result is obtained by extrapolating to $\epsilon=0$, see Fig.~2.

This protocol relies
on the following crucial mathematical assumptions: 
{\it (i)} the $N$-th order contribution of the diagrammatic expansion for
$\Sigma$ 
(for fixed external variables)
is the $N$-th coefficient
of the Taylor series at $z=0$ of a function $g(z)$   
which has a non-zero convergence radius, 
 {\it (ii)} the analytic continuation $g(1)$,
performed by the above resummation methods~\cite{Hardy49,fru92resum},
is the physically correct value of
$\Sigma$.
The same assumptions should hold for $\Pi$.

Proving these assumptions is an open mathematical challenge.
Note that Dyson's collapse argument~\cite{dys52} is not applicable to immediately disprove
the assumption~{\it (i)} of a non-zero convergence radius: Indeed, unlike QED, our skeleton series is not an expansion in powers of a coupling constant whose sign change would lead to an instability.
A first important evidence for the validity of our mathematical assumptions is that the three different resummation methods yield consistent results.
For an independent test, we turn to experiments.

\begin{figure}
    \begin{center}
    \includegraphics[width= \columnwidth]{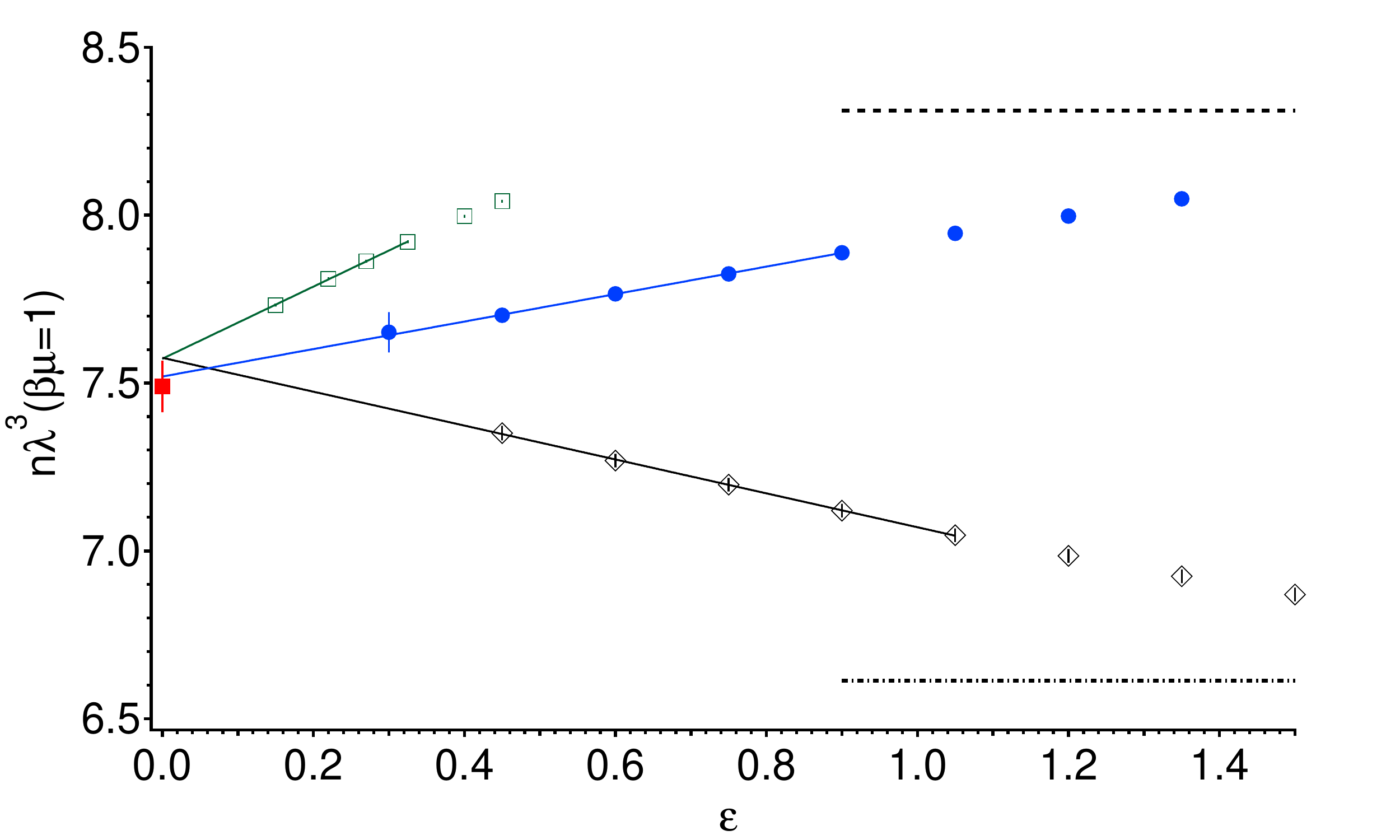}
    \caption[Title]{{\bf Cross-validation between resummation procedure
    and experiment at $\beta\mu = +1$.}
 Bold Diagrammatic Monte Carlo data for the dimensionless density $n\lambda^3$, as a function of the parameter $\epsilon$ controlling the resummation procedure,
 for
 three different resummation methods:
 Lindel\"{o}f
 (blue circles), shifted Lindel\"{o}f
 (black diamonds), and Gauss (open green squares).
  The solid lines are linear fits to the Monte Carlo data, their $\epsilon\rightarrow 0$ extrapolation agrees within error bars
  with the experimental data point
   (filled red square).
[In the opposite limit $\epsilon\to\infty$, the
Lindel\"{o}f (resp. shifted Lindel\"{o}f) curves will asymptote to
the first~\cite{haus94,haus07bcsbec} (resp. third) order results, shown by
the dashed (resp. dash-dotted) line.]
Error bars for each $\epsilon$ represent the statistical error, together with the estimated systematic error coming from not sampling diagrams of order $>9$. }
    \label{fig:resummation}
    \end{center}
\end{figure}

The present experiment furnishes high-precision data for the density $n$ as a function of the local value $V$ of the trapping potential (see Fig.~3 and Methods).
We start the process by obtaining the EOS at high temperatures
in the non-degenerate wings of the atom cloud where the virial expansion is applicable.
Once the temperature and the chemical potential 
have been determined from fits to the wings of the cloud,
the data closer to the cloud center provides a new prediction of the EOS.
The process is iterated to access lower temperatures.

\begin{figure}
    \begin{center}
    \includegraphics[width= \columnwidth]{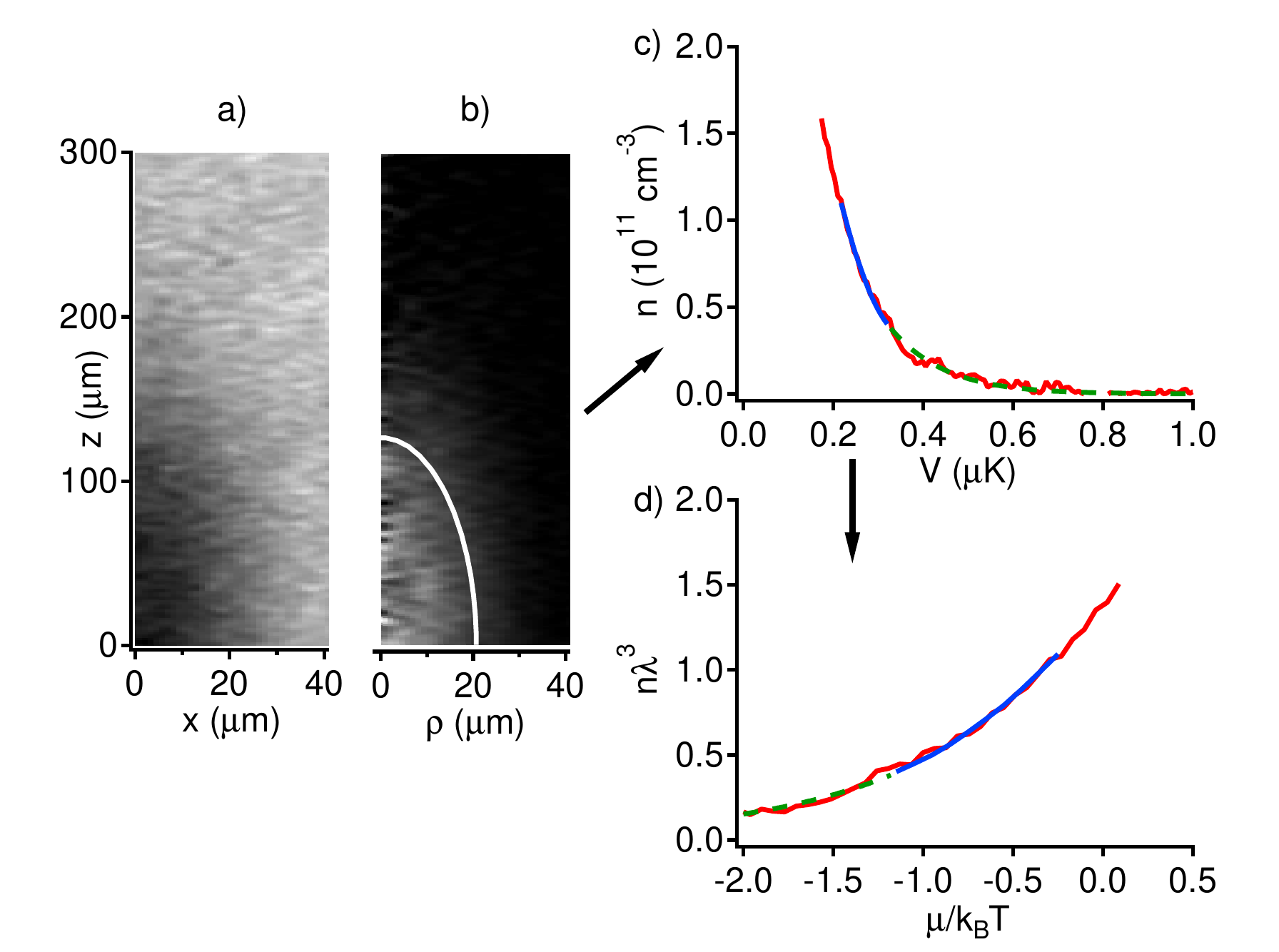}
    \caption[Title]{{\bf Constructing the EOS from {\it in situ} imaging.} The atom cloud shown contains $N=8 \times10^4$ atoms for each spin-state, with a local Fermi energy of $E_F=370\,\rm n$K at the center. a) 
    Absorption image of the atomic cloud after quadrant averaging.
    b) Reconstructed local density $n(\rho,z)$. c) Equipotential averaging produces a low-noise density profile $n$ vs $V$. Thermometry is performed by fitting the experimental data (red) to the known portion of the EOS (solid blue line), starting with the virial expansion for $\beta\mu<-1.25$ (green dashed line). In this example, the EOS is known to $\beta\mu\leq-0.25$, and the fit to the density profile yields $T=113\,n$K, and $\beta\mu=1.63$. d) Given $\mu$ and $T$, the density profile can be rescaled to produce the EOS $n\lambda^3$ vs $\beta\mu$.}
    \label{fig:experiment}
    \end{center}
\end{figure}


Scale invariance allows to write the density EOS 
as $n(\mu,T)\lambda^3 = f(\beta\mu)$, with 
$\lambda=\sqrt{2\pi\hbar^2/(m k_B T)}$ the thermal de~Broglie wavelength, $\beta=1/(k_B T)$ the inverse temperature and $f$ a universal function. A convenient normalization of the data is provided by the EOS of a non-interacting Fermi gas, $n_0 \lambda^3 = f_0(\beta\mu)$.
 In Fig.~4a we thus report the ratio $n(\mu,T)/n_0(\mu,T)= f(\beta\mu)/f_0(\beta\mu)$, bringing out the difference between the ideal and the strongly interacting Fermi gas.
The Gibbs-Duhem relation allows us to also calculate the pressure at a given chemical potential,
$P(\mu_0,T) = \int_{-\infty}^{\mu_0} {\rm d}\mu\; n(\mu,T) = \frac{1}{\beta \lambda^3} F(\beta\mu_0)$,
where $F(x) = \int_{-\infty}^x {\rm d}x' f(x')$. We normalize it by the pressure of
the ideal Fermi gas and show $F(\beta\mu)/F_0(\beta\mu)$, see Fig.~4b. 
The agreement between BDMC and experiment is excellent.
The comparison is sufficiently sensitive to validate the procedure of resumming and extrapolating (see Fig.~2). 
The result was checked to be independent of the
 maximal sampled diagram order $N_{\rm max}\in\{7;8;9\}$
 within the error bars displayed in Fig.~2 for each $\epsilon$.
The BDMC final error bar in Fig.~4 is the sum of the conservatively estimated systematic errors from the uncertainty of the $\epsilon\to0$ extrapolation and from the dependence on numerical grids and cutoffs, the latter being reduced by analytically treating high-momentum short-time singular parts.
The systematic
error in the experiment is 
determined to be about 1\%
by the
independent determination of the EOS of the non-interacting Fermi gas.
The experimental error bars of Fig. 4 also include the statistical error, which is
$<0.5\%$ thanks to the
scale invariance of the balanced unitary gas:
irrespective of shot-to-shot fluctuations of atom number and temperature,
all
experimental profiles contribute to the same scaled EOS-function $f$. 
The dominant 
uncertainty on
 the experimental EOS stems from the uncertainty in the position of the $^6$Li Feshbach resonance known to be at $834.15 \pm 1.5 \,\rm G$ from spectroscopic measurements~\cite{bart04fesh}.
The change in energy, pressure and density with respect to the interaction
strength is controlled by the so-called contact~\cite{braa11lect}
that is obtained from $\Gamma$ in the BDMC calculation. 
This allows us to define the uncertainty margins above and below the experimental data (see Fig.~4) that give the prediction for the unitary EOS if the true Feshbach resonance lied 1.5 G below or above 834.15 G, respectively.

\begin{figure*}[ht]
    \begin{center}
    \includegraphics[width= \linewidth]{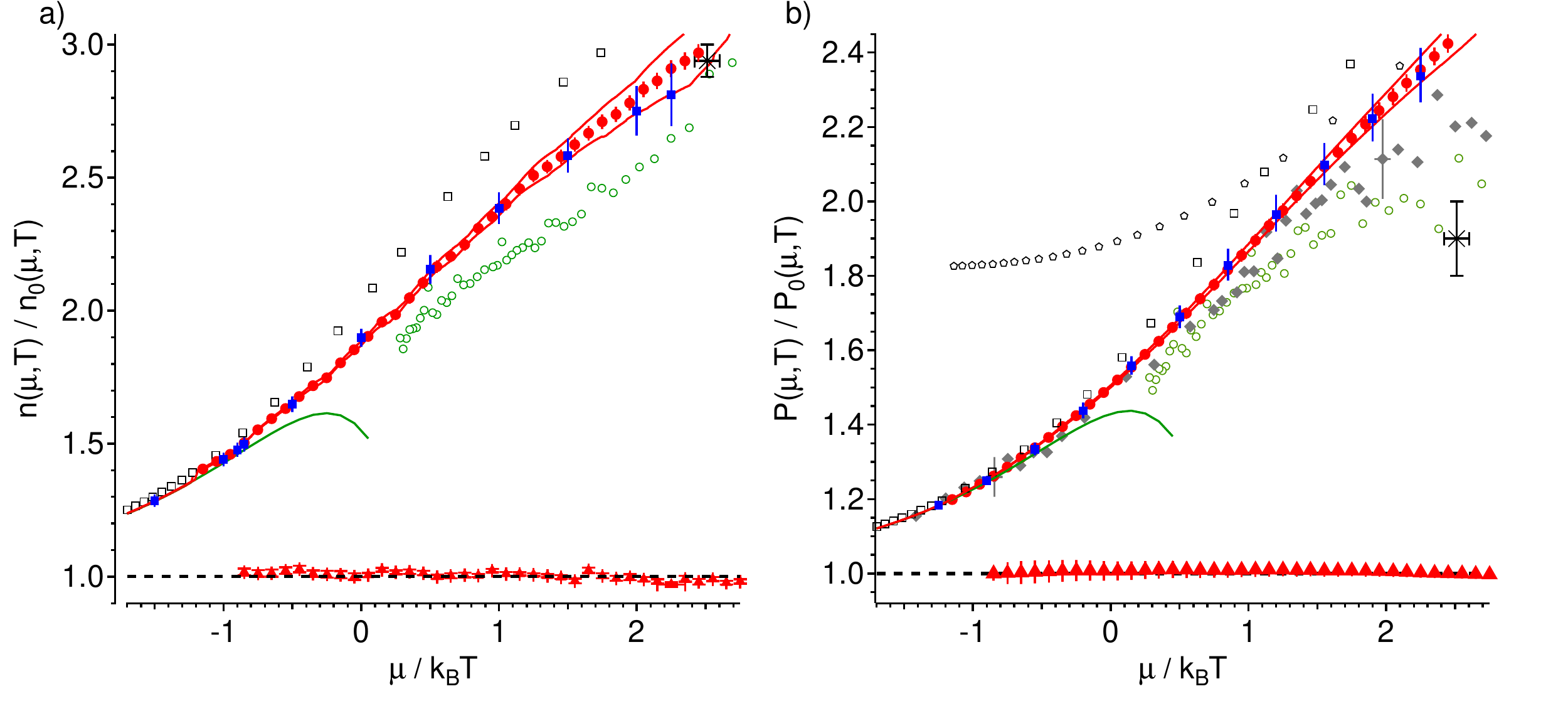}
    \caption[width= 2\columnwidth]{{\bf Equation of state of the unitary Fermi gas in the normal phase.} {\bf (a)} Density $n$ and {\bf (b)} pressure $P$ of a unitary Fermi gas, 
    normalized by the density $n_0$ and the pressure $P_0$ of a non-interacting Fermi gas,
    versus the ratio of chemical potential $\mu$ to temparature $T$. 
 Blue filled squares: Bold Diagrammatic Monte Carlo (this work),
red filled circles: MIT experiment (this work).
The BDMC error bars are estimated upper bounds on systematic errors.
The MIT error bars are one standard deviation systematic plus statistical errors, with the additional uncertainty from the Feshbach resonance position shown by the upper and lower margins in red solid lines.
    Black dashed line and red triangles: Theory and MIT experiment (this work) for the ideal Fermi gas,
    used to assess the experimental systematic error.
     Green solid line: third order virial expansion.
      Open squares: first order bold diagram~\cite{haus94,haus07bcsbec}.
       Green open circles: Auxiliary Field QMC~\cite{bulg06TC}.
        Star: superfluid transition point from Determinental Diagrammatic Monte Carlo~\cite{goul10tc}.
         Filled diamonds: ENS experimental pressure EOS~\cite{nasc10thermo}.
          Open pentagons: pressure EOS from Tokyo experiment~\cite{hori10thermo}.
}
    \label{fig:eos}
    \end{center}
\end{figure*}

We clearly discriminate against previous theoretical and experimental results.
Deviations
from the theory based on the
first-order Feynman diagrams~\cite{haus94,haus07bcsbec} are expected, and rather remarkably moderate. 
Differences with
 lattice Monte Carlo data~\cite{bulg06TC,goul10tc} may seem more surprising,
since in the particular case of the balanced system these algorithms are
free of the sign problem, allowing in principle to approach the balanced unitary gas model in an unbiased way.
However, eliminating systematic errors from lattice-discretization and finite volume
requires extrapolations which are either not done~\cite{bulg06TC}
or difficult to control~\cite{buro06TC,goul10tc}. 
The ENS experimental pressure EOS~\cite{nasc10thermo}  lies systematically below ours, slightly
outside the reported error bar.
The experimental results from Tokyo~\cite{hori10thermo} do not agree with the virial expansion
at high temperature.
The BDMC results agree excellently with the present experimental data all the way
down to the critical temperature for superfluidity (see Fig.~4).
On approach to $(\beta \mu)_c$, we observe the growth of the correlation length in the BDMC pair correlation function $\Gamma$.
A protocol for extracting the critical temperature itself from the BDMC simulation
will be presented elsewhere.

We are not aware of any system of strongly correlated fermions in Nature where experimental and unbiased theoretical results were compared at the same level of accuracy.
Even for bosons, the only analog is liquid $^4$He.
This promotes the unitary gas to the major testing ground for unbiased theoretical treatments. 
The present BDMC implementation should remain applicable at finite polarization and/or finite scattering length, opening the way to rich physics which was already addressed by cold atoms experiments~\cite{kett08rivista,zwer11book,shin07phasediagram,shin08eos,navo10eos}. We also plan to extend BDMC to superfluid phases by introducing anomalous propagators.
Moreover, since the method is generic, we expect
 numerous other important applications to
long-standing problems across many fields.

\noindent{\it Note added in proof}: After a preprint of this work became available, new Auxiliary Field QMC data were presented~\cite{drut11update}, with undetermined systematic errors whose evaluation in future work is called for by the authors of Ref.~\cite{drut11update}.

\vskip 2cm

\textbf{Methods}

The experimental setup has been described previously~\cite{kett08rivista}. In short, ultracold fermionic $^6$Li is brought to degeneracy via sympathetic cooling with $^{23}$Na. A two-state mixture of the two lowest hyperfine states of $^6$Li is further cooled in a hybrid magnetic and optical trap at the broad Feshbach resonance at 834 G.
We employ high-resolution {\it in situ} absorption imaging to obtain the column density of the gas, that is converted into the full 3D density via the inverse Abel transform~\cite{shin06phase}. Equidensity lines provide equipotential lines that are precisely calibrated via the known axial, harmonic potential (axial frequency $\nu_z=22.83\pm 0.05$ Hz). Equipotential averaging yields low-noise profiles of density $n$ versus potential $V$. Density is absolutely calibrated by imaging a highly degenerate, highly imbalanced Fermi mixture, and fitting the majority density profile to the ideal Fermi gas EOS~\cite{kett08rivista}. In contrast to previous studies~\cite{nasc10thermo,hori10thermo}, our analysis does not rely on the assumption of harmonic trapping.

Thermometry is performed by fitting the density profile to the EOS constructed thus far, restricting the fit to the portion of the density profile where the EOS is valid.
In the high-temperature regime, the EOS is given by the virial expansion
\begin{equation}
n\lambda^3=e^{\beta\mu}+2 b_2 e^{2\beta\mu}+3 b_3 e^{3\beta\mu}+...\,
\label{eq:virial3}
\end{equation}
where the virial coefficients are $b_2=3\sqrt{2}/8$~\cite{ho04virial}, and $b_3=-0.29095295$~\cite{liu09virial}. 
Fitting a high-temperature cloud to the virial expansion gives the temperature $T$ and the chemical potential $\mu$ of the cloud, and the EOS $n\lambda^3=f(\beta\mu)$ can be constructed. 
We have used Eq.(\ref{eq:virial3}) for $\beta\mu<(\beta\mu)_{max}=-1.25$ and we checked that our EOS did not change within statistical noise if we instead used $(\beta\mu)_{max}=-0.85$.
Once a new patch of EOS has been produced, it can then in turn be used to fit colder clouds. Iteration of this method allows us to construct the EOS to arbitrarily low temperature.
A total of $\sim 1000$ profiles were used, with 10 to 100 profiles (50 on average) contributing
at any given $\beta\mu$.


\textbf{Acknowledgements}

We thank R. Haussmann for providing propagator data from~\cite{haus94,haus07bcsbec} for comparison, and the authors of Refs.~\cite{goul10tc,bulg06TC, hori10thermo,nasc10thermo} for sending us their data.
This collaboration was supported by a grant from the Army Research Office with funding from the DARPA Optical Lattice Emulator program.
 Theorists 
acknowledge the financial support of the Research Foundation - Flanders FWO (K.V.H.), 
 NSF grant 
 PHY-1005543 (UMass group), SNF Fellowship for Advanced Researchers (E.K.), and IFRAF (F.W.).
 Simulations ran on the clusters CM at UMass and brutus at ETH.
 The MIT work was supported by the NSF, AFOSR-MURI, ARO-MURI, ONR, DARPA YFA, an AFOSR PECASE, the David and Lucile Packard Foundation, and the Alfred P. Sloan Foundation.

\textbf{Author Contributions}

  K.V.H. (theory) and M.J.H.K. (experiment) contributed equally to this work.
K.V.H., F.W., E.K., N.P. and B.S. developed the BDMC approach for unitary fermions; the computer code was written by K.V.H. assisted by F.W.; simulation data were produced by F.W., E.K. and K.V.H.; M.J.H.K., A.T.S., L.W.C., A.S. and M.W.Z. all contributed to the experimental work and the analysis. 
All authors participated in the manuscript preparation.

\textbf{Additional Information}

Correspondence and requests for materials
should be addressed to K.V.H.~(email:kris.vanhoucke@ugent.be).

\bibliography{References}

\end{document}